\documentclass[10pt, twocolumn, conference]{IEEEtran}
\IEEEoverridecommandlockouts
\usepackage{geometry}
\usepackage[nospace,noadjust]{cite}
\usepackage{amsmath,amssymb,amsfonts,times,slashbox}
\usepackage{graphicx}
\usepackage{textcomp}
\usepackage{xcolor}
\usepackage{subfigure}
\usepackage{epsfig}
\usepackage{multirow}
\usepackage{algorithm}
\usepackage{algorithmicx}
\usepackage{algpseudocode}
\usepackage{array}
\usepackage{epstopdf}
\usepackage{color}
\usepackage{diagbox}

\title{\LARGE \bf Dynamic Compressive Sensing based on RLS for Underwater Acoustic Communications}

\author{{\normalsize Zhen Qin$^{\dag}$}
\vspace{0.1cm}
\\{\normalsize $^\dag$Department of Computer Science and Engineering, Ohio State University, Columbus, OH 43210, USA}}

\begin{document}
\maketitle

\begin{abstract}


Sparse structures are widely recognized and utilized in channel estimation.
Two typical mechanisms, namely proportionate updating (PU) and zero-attracting (ZA) techniques, achieve better performance, but their computational complexity are higher than non-sparse counterparts.
In this paper, we propose a DCS technique based on the recursive least squares (RLS) algorithm which can simultaneously achieve improved performance and reduced computational complexity. Specifically, we develop the sparse adaptive subspace pursuit-improved RLS (SpAdSP-IRLS) algorithm by updating only the sparse structure in the IRLS to track significant coefficients. The complexity of the SpAdSP-IRLS algorithm is successfully reduced to $\mathcal{O}(L^2+2L(s+1)+10s)$, compared with the order of $\mathcal{O}(3L^2+4L)$ for the standard RLS. Here, $L$ represents the length of the channel, and $s$ represents the size of the support set. Our experiments on both synthetic and real data show the superiority of the proposed SpAdSP-IRLS, even though only $s$ elements are updated in the channel estimation.

%
\end{abstract}

\section{Introduction}
\label{sec:intro}

The sparse structure of the channel, such as the digital TV transmission and underwater acoustic (UWA) communications, has been widely recognized \cite{cotter2002sparse,Li07,Berger09,Pelekanakis13,Pelekanakis14,Duan18,Tao19}, motivating the design of sparse adaptive filtering algorithms \cite{Chen09,Gu09,Eksioglu111,Qin19,Duttweiler00,Benesty02,Loganathan09,Qin20,QinDSP22}.
The proportionate updating (PU) and zero-attracting (ZA) principles have been mainly utilized in the existing algorithms.
The ZA-type algorithms were developed by adopting a regularized cost function to attract inactive taps to zero in the strictly sparse system \cite{Chen09,Gu09,Eksioglu111,Qin19},
while, for systems which are not strictly sparse but possess a relatively sparse (nonuniform) structure, the PU-type algorithms achieve a better performance  \cite{Duttweiler00,Benesty02,Loganathan09,Qin20,QinDSP22}. With the advancement in the sparse adaptive filtering algorithms, the investigation of the channel estimation in the UWA communications resurges \cite{Pelekanakis10,Wu13,Tian19}.

However, compared with non-sparse counterparts, sparse adaptive algorithms with two mechanisms mentioned above achieve an improved performance at the cost of high complexity. The reason behind this is that algorithms still need to update all elements rather than a support set. Recently, there has been significant interest in researching dynamic compressive sensing (DCS) techniques that track only the positions of significant taps and update them using sparsity. Different from the traditional compressive sensing (CS) methods, the DCS allows one adjustable batch mode, and hence is suitable for time-varying environments.

The algorithms for the DCS can be classified into three categories: dynamic convex optimization methods \cite{Vaswani08}, the dynamic approximate Bayesian approach \cite{Ziniel13}, and dynamic greedy algorithms \cite{Mileounis10,Vlachos12,Zachariah12,Qin202,qin2022adaptive}.
Compared to the first two methods, dynamic greedy algorithms have lower computational complexity, making them more suitable for online operation. In \cite{Mileounis10},  the inherent batch-mode compressive sampling matching pursuit (CoSaMP) method was successfully incorporated into the DCS algorithm, resulting in the sparse adaptive orthogonal matching pursuit (SpAdOMP) method. To be specific, the SpAdOMP method replaced the least squares (LS) algorithm in CoSaMP with the normalized least mean square (NLMS) adaptive filter algorithm.
However, colored input signals can significantly degrade its convergence speed.
Recently, we proposed the SpAdOMP-affine projection algorithm (SpAdOMP-APA) \cite{Qin202} and sparse adaptive subspace pursuit-improved proportionate APA (SpAdSPP-IPAPA) \cite{qin2022adaptive}. Compared with the NLMS-type greedy algorithms, they achieve a faster convergence speed and a better performance.

Although first-order adaptive algorithms, such as NLMS and APA, have been extensively studied, their linear convergence speed is often slow in practice, necessitating a larger amount of training data. To further improve the convergence rate of the dynamic greedy algorithm, we will explore a typical second-order adaptive algorithm, the recursive least squares (RLS) algorithm, whose convergence rate is super-linear.
In this paper, we first propose the sparse adaptive subspace pursuit-improved recursive least squares (SpAdSP-IRLS) algorithm whose complexity is lower than the standard RLS algorithm.   When the length of the channel is $L$ and the size of the support set $s$, the complexity of the SpAdSP-IRLS is $\mathcal{O}(L^2+2L(s+1)+10s)$ instead of $\mathcal{O}(3L^2+4L)$ for the standard RLS. Moreover, the proposed SpAdSP-IRLS can achieve improved performance since updating only the support set can further increase the convergence rate and reduce the noise in the sparse weights. Experimental results corroborated the superiority of the SpAdSP-IRLS in the UWA channel estimation.

{\bf Notation}: Throughout the paper, we use bold uppercase (lowercase) letters to denote matrix, $\bf{A}$, and vector, $\bf{a}$, respectively. An italic letter, $a$, represents a scalar quantity. The superscripts, $(\cdot)^T$, $(\cdot)^*$, and $(\cdot)^H$ denote the transpose, complex conjugate, and Hermitian transpose operators, respectively.
The ${\rm supp}({\bf a},s)$ denotes the support set corresponding to the $s$ largest components in vector ${\bf a}$.
${\rm Im}(a)$ takes the imaginary part of a complex value $a$.
The ${\bf I}_L$ is an identity matrix of size $L$.

\section{Dynamic Compressive Sensing Based Channel Estimation}
\label{DCSBCE}

For the UWA communication, the received signal at time n is given as
\begin{equation}
\label{DCSBCEl}
y(n)=e^{j \theta(n)}\sum_{l=0}^{L-1}h_l(n)x(n-l)+q(n),
\end{equation}
where $h_l(n)$ is the $l$-th channel tap, $\theta(n)$ is the phase rotation accounting for the Doppler effect of the UWA channel \cite{TaoOceans19}, $x(n-l)$ is the transmitted symbol, and $q(n)$ denotes an additive noise. Define the channel vector at time $n$ as ${\bf h}(n)=[h_0(n),...,h_{L-1}(n)]^T$ and $\textbf{x}(n)=[x(n), x(n-1), ... ,x(n-L+1)]^T$.

\subsection{Dynamic Tracking of the Support Set}
\label{SpIPAPA_support}

Next, we will detail the dynamic tracking of SpAdSP-IRLS algorithm.
Define the proxy signals for the proposed algorithm as
\begin{eqnarray}
    \label{DCSBCE2}
    {\bf p}(n)=\eta{\bf p}(n-1)+e^{-j\widehat{\theta}(n)}{\bf x}^*(n)v(n-1),
\end{eqnarray}
where $\eta\in(0,1]$ is a forgetting factor. In time-invariant environments, $\eta$ shall be set to 1.
The $\widehat{\theta}(n)$ is an estimation of the phase rotation that can be produced by a second-order phase locked loop (PLL) \cite{stojanovic94}.
The $v(n-1)$ as the symbol estimation error at time $n-1$ is denoted as
\begin{eqnarray}
    \label{DCSBCE3}
    v(n-1)=y(n-1)-e^{j\widehat{\theta}(n-1)}\widetilde{{\bf h}}^H(n-1){\bf x}(n-1),
\end{eqnarray}
where $ \widetilde {\bf{h}}(n-1)$ is the applied estimator vector and obtained from intermediate estimator vector $\widehat{{\bf h}}(n)$ as
\begin{eqnarray}
    \label{DCSBCE4}
    {\widetilde{{\bf h}}}_{\Lambda_{s}}(n-1)=\widehat{{\bf h}}_{\Lambda_{s}}(n-1),\ \ {\widetilde{{\bf h}}}_{\Lambda^c_{s}}(n-1)={\bf 0},
\end{eqnarray}
where ${\Lambda_{s}}$ of size $s$ is the support set at time $n-1$. Accordingly, its complementary set ${\Lambda^{c}_{s}}$ consists of $L-s$ elements. For brevity, time indices are dropped from the notations of all four sets.

Based on ${\bf p}(n)$, the support set $\Lambda$ for intermediate estimator vector, $\widehat{{\bf h}}(n)$, is updated as
\begin{eqnarray}
    \label{DCSBCE5}
    \Lambda=\Omega\cup{\rm supp}({\widetilde{{\bf h}}}(n-1),s),
\end{eqnarray}
where according to the SP algorithm \cite{Dai09}, $\Omega={\rm supp}({\bf p}(n),s)$ represents the support sets corresponding to the $s$ largest components in ${\bf p}(n)$. Again, the time index $n$ has been dropped in $\Lambda$ and $\Omega$ for notation simplicity.

\subsection{Adaptive Updating of the Coefficients}
\label{AUC}

In the SpAdSP-IRLS algorithm, the IRLS algorithm is adopted to track the estimator coefficients.
In order to distinguish it from the standard RLS algorithm, we note that  the step size is introduced in the updating equation of the IRLS in order to improve performance in the practical implementation.

Define the apriori error as
\begin{eqnarray}
    \label{AUC1}
    e(n|n-1)=y(n)-e^{j\widehat{\theta}(n)}{\widehat{{\bf h}}}_{\Lambda}^H(n-1){\bf x}_{\Lambda}(n),
\end{eqnarray}
where ${\widehat{{\bf h}}}_{\Lambda}(n-1)\in\mathcal{C}^{|\Lambda|\times 1} $ and ${\bf x}_{\Lambda}(n)\in\mathcal{C}^{|\Lambda|\times 1}$ are respectively submatrices of ${\widehat{{\bf h}}}(n-1)$ and ${\bf x}(n)$.
The updating of the estimator vector over its support sets is then
\begin{eqnarray}
    \label{AUC2}
    {\widehat{{\bf h}}}_{\Lambda}(n)={\widehat{{\bf h}}}_{\Lambda}(n-1)+\mu{\bf k}_{\Lambda}(n)e^*(n|n-1),
\end{eqnarray}
where $\mu$ is the step size that introduced in the updating equation of the traditional RLS algorithm. Compared with the standard RLS algorithm, the step size $\mu$ is introduced in order to better balance the convergence and steady-state error. In addition, the Kalman gain vector ${\bf k}_{\Lambda}(n)\in\mathcal{C}^{|\Lambda|\times 1} $ is the subvector of ${\bf k}(n)$ corresponding to the support set $\Lambda$, given by
\begin{eqnarray}
    \label{AUC3}
    {\bf k}(n)=\frac{e^{j\widehat{\theta}(n)}{\bf P}(n-1){\bf x}(n,\Lambda)}{\lambda+{\bf x}^H(n,\Lambda){\bf P}(n-1){\bf x}(n,\Lambda)},
\end{eqnarray}
with $\lambda$ denoting  a forgetting factor ranging in $(0,1]$. ${\bf x}(n,\Lambda)={\bf Q}(n,\Lambda){\bf x}(n)\in\mathcal{C}^{L\times 1}$, where the diagonal matrix ${\bf Q}(n,\Lambda)$ can be designed by ${Q}_{i}(n,\Lambda)=\begin{cases}
    1, & i\in\Lambda\\
    0, & i\notin\Lambda
    \end{cases}$,
denotes that elements of ${\bf x}(n,\Lambda)$ in the support set $\Lambda$ are same with ${\bf x}(n)$ and the rest of them are $0$.
The matrix ${\bf P}(n-1)$ is the inverse of the input covariance matrix, which can be iteratively computed as
\begin{eqnarray}
    \label{AUC4}
    {\bf P}(n-1)&\!\!=\!\!&\lambda^{-1}\bigg[{\bf P}(n-2)-e^{-j\widehat{\theta}(n-1)}{\bf k}(n-1)\nonumber\\
    &&\cdot\bigg({\bf x}^H(n-1,\Lambda){\bf P}(n-2)\bigg)\bigg],
\end{eqnarray}

Based on ${\widehat{{\bf h}}}_{\Lambda}(n)$ in \eqref{AUC2}, we can obtain
\begin{eqnarray}
    \label{AUC5}
    \Lambda_{s}&={\rm supp}({\widehat{{\bf h}}}_{\Lambda}(n),s),
\end{eqnarray}
which is required in \eqref{DCSBCE4}.

In the end, the phase updating of the second-order PLL is
\begin{eqnarray}
    \label{AUC6}
    {\widehat{\theta}(n)}=\widehat{\theta}(n-1)+K_1\phi(n-1)+K_2\sum_{t=0}^{n-1}\phi(t),
\end{eqnarray}
where $\phi(n-1)={\rm Im}(y^*(n-1)e^{j\widehat{\theta}(n-1)}{\widehat{{\bf h}}}^H(n-1){\bf x}(n-1))$.

The implementation of the SpAdSP-IRLS is finally summarized in Algorithm 1.
Last, a complexity comparison between RLS and SpAdSP-IRLS algorithms in terms of complex multiplication (CM) is shown in Table~\ref{SpIPAPA_CM_1}.
Clearly, the comparison between two algorithms depends on the choice of $s$. Moreover, we note that the complexity $L^2$ in the SpAdSP-IRLS is from \eqref{AUC4} since ${\bf k}(n-1)$ and ${\bf x}^H(n-1,\Lambda){\bf P}(n-2)$ are full vectors without using support set. We leave the exploration of lower-complexity methods for future work.



\floatname{algorithm}{Algorithm}
\renewcommand{\algorithmicrequire}{\textbf{Initialization:}}
\renewcommand{\algorithmicensure}{\textbf{Iteration:}}
\begin{algorithm}
\begin{algorithmic}
\caption{SpAdSP-IRLS with Second-order PLL}
\Require ${\widetilde{{\bf h}}}(0)=[0,0,...,0]^T\in\mathcal{C}^{L\times 1}$,  ${\widehat{{\bf h}}}_{\Lambda}(0)=[0,0,...,0]^T\in\mathcal{C}^{|\Lambda|\times 1}$ with $|\Lambda|=s$ at time $0$, ${\bf p}(0)=[0,0,...,0]^T\in\mathcal{C}^{L\times 1}$, $v(0)=y(1)$, $\widehat{\theta}(0)=0$ and ${\bf P}(0)=\delta^{-1}{\bf I}_L\in\mathcal{C}^{L\times L}$. The $\mu$, $K_{1}$, $K_{2}$, $\lambda$, $\delta$, $\eta$ and $s$ are predetermined;
\Ensure \\
\ \hspace{-0.3cm}For $n=1,2,\cdots$ \\
\ ${\bf p}(n)=\eta{\bf p}(n-1)+e^{-j\widehat{\theta}(n)}{\bf x}^*(n)v(n-1),$\\
\ $\Omega={\rm supp}({\bf p}(n),s),$\\
\ $\Lambda=\Omega\cup{\rm supp}({\widetilde{{\bf h}}}(n-1),s),$\\
\ ${\bf Q}(n,\Lambda)={\rm diag}\{{Q}_{1}(n,\Lambda),\dots,{Q}_{L}(n,\Lambda)\},$ where\\
\ ${Q}_{i}(n,\Lambda)=\begin{cases}
    1, & i\in\Lambda\\
    0, & i\notin\Lambda
    \end{cases}$,\\
\ ${\bf x}(n,\Lambda)={\bf Q}(n,\Lambda){\bf x}(n),$\\
\ ${\bf k}(n)=\frac{e^{j\widehat{\theta}(n)}{\bf P}(n-1){\bf x}(n,\Lambda)}{\lambda+{\bf x}^H(n,\Lambda){\bf P}(n-1){\bf x}(n,\Lambda)},$\\
\ $e(n|n-1)=y(n)-e^{j\widehat{\theta}(n)}{\widehat{{\bf h}}}_{\Lambda}^H(n-1){\bf x}_{\Lambda}(n),$\\
\ ${\widehat{{\bf h}}}_{\Lambda}(n)={\widehat{{\bf h}}}_{\Lambda}(n-1)+\mu{\bf k}_{\Lambda}(n)e^*(n|n-1),$\\
\ ${\bf P}(n)=\lambda^{-1}[{\bf P}(n-1)-e^{-j\widehat{\theta}(n)}{\bf k}(n)({\bf x}^H(n,\Lambda){\bf P}(n-1))],$\\
\ $\Lambda_{s}={\rm supp}({\widehat{{\bf h}}}_{\Lambda}(n),s),$\\
\ ${\widetilde{{\bf h}}}_{\Lambda_{s}}(n-1)=\widehat{{\bf h}}_{\Lambda_{s}}(n-1),\ \ {\widetilde{{\bf h}}}_{\Lambda^c_{s}}(n-1)={\bf 0},$\\
\ $v(n)=y(n)-e^{j\widehat{\theta}(n)}\widetilde{{\bf h}}^H(n){\bf x}(n),$\\
\ ${\widehat{\theta}(n)}=\widehat{\theta}(n-1)+K_1\phi(n-1)+K_2\sum_{t=0}^{n-1}\phi(t)$, where\\
\ $\phi(n-1)={\rm Im}(y^*(n-1)e^{j\widehat{\theta}(n-1)}{\widehat{{\bf h}}}^H(n-1){\bf x}(n-1)).$\\
\ \hspace{-0.3cm}end
\end{algorithmic}
\end{algorithm}

\begin{table}[!ht]
\renewcommand{\arraystretch}{1.5}
\begin{center}
\caption{The complexity comparison between RLS and SpAdSP-IRLS} 
\label{SpIPAPA_CM_1}
{\begin{tabular}{|c||p{5cm}<{\centering}|}\hline  {\bf Algorithms} & {\bf Complexity (in terms of CMs)}
\\\hline {\bf RLS} & $\mathcal{O}(3L^2+4L)$
\\\hline {\bf SpAdSP-IRLS} & $\mathcal{O}(L^2+2L(s+1)+10s)$\\\hline
\end{tabular}}{}
\end{center}
\end{table}

\section{Experimental Results}
\label{ExperiResults}

To verify the performance gain of the proposed SpAdSP-IRLS algorithm, both simulated and experimental examples of system identification were conducted. Before presenting experimental results, we first define two algorithms. To be specific, (1) the SpAdSP-IRLS becomes IRLS when $s=L$; (2) the SpAdSP-IRLS reduces to SpAdSP-RLS when $\mu=1$.

\subsection{Synthetical data}
\label{SD}

\begin{figure}[!ht] 
   \hspace{0.5cm}
   \includegraphics[width=7.5cm,keepaspectratio]
   {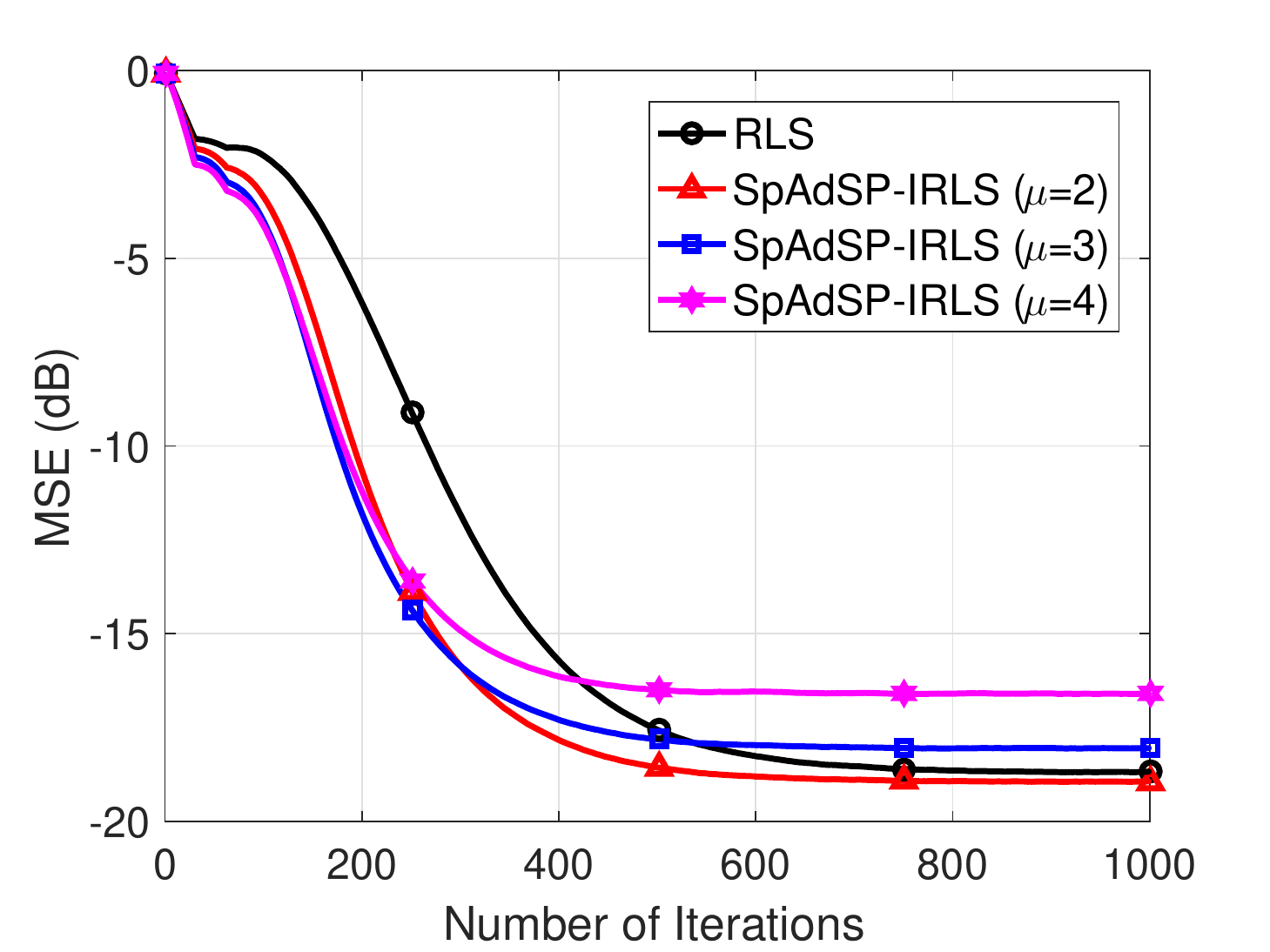}
   \caption{Performance comparison between RLS and SpAdSP-IRLS.}
   \label{Figure1}
\end{figure}

In the first experiment, we considered a sparse system of length $N = 64$, in which the system coefficients with indices $1, 32, 33$ and $64$ were $0.3536 + 0.3536i$.
The system input, $x(n)$, was a zero-mean white Gaussian process with variance $\sigma_x^2=1$.
The additive noise $v(n)$ was also zero-mean white Gaussian with its variance determined by the signal to noise ratio (SNR), where ${\rm SNR}=20{\rm dB}$.
For all considered algorithms including the RLS and the proposed SpAdSP-IRLS, the forgetting factor is $\lambda=0.99$ and $\delta=100$.
For the SpAdSP-IRLS, $\eta=0.5$ and $s=12$.
The results are shown in Fig.~\ref{Figure1}, where all curves were obtained by averaging over 1000 independent trials.
Obviously, the proposed SpAdSP-IRLS algorithm outperforms the RLS even though the size of the support set is only $12$.
In addition, we can notice that the smaller is $\mu$, the better steady-state performance is achieved for the SpAdSP-IRLS algorithm over the RLS algorithm.

\begin{figure}[!ht] 
   \hspace{0.5cm}
   \includegraphics[width=7.5cm,keepaspectratio]
   {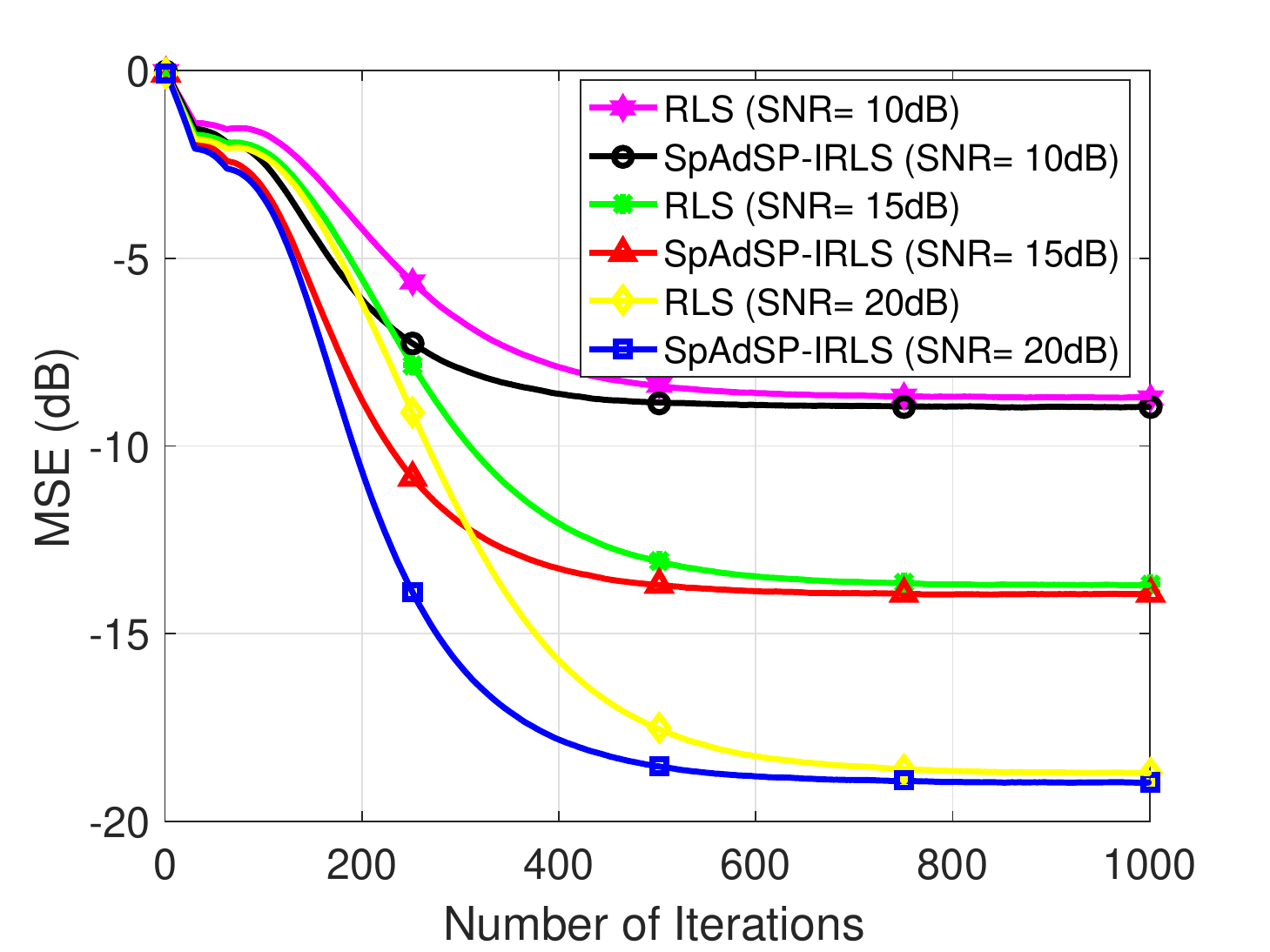}
   \caption{Performance comparison between RLS and SpAdSP-IRLS under different SNRs ($\mu=2$).}
   \label{Figure2}
\end{figure}

In the second experiment, the performance of the RLS and the proposed SpAdSP-IRLS under different SNRs have been shown in Fig.~\ref{Figure2}.
The SpAdSP-IRLS achieved a better performance than the RLS in terms of convergence and steady-state error for different SNRs.
It also showed the parameter choice is not sensitive to SNR.

\subsection{Real data}
\label{RD}

In this subsection, we will testify the performance the proposed SpAdSP-IRLS via real-world experimental data collected in a shallow-water acoustic communication trial.
The experiment was performed at the coastline of Xiamen, China in 2019. The carrier frequency was 16 kHz and the symbol period was 0.25 ms.
The RLS, IRLS, SpAdSP-RLS and SpAdSP-IRLS were employed to achieve adaptive channel estimation.
For all algorithms, $\lambda=0.99$ and $\delta=100$ and for SpAdSP-type algorithms, $\eta=0.999$ and $s=15$.
For IRLS and SPAdSP-IRLS, $\mu=1.6$.
The parameters for the second order PLL were: $K_1=-5\times 10^{-12}$ and $K_2=K_1/10$.

\begin{figure}[!ht] 
   \hspace{0.5cm}
   \includegraphics[width=7.5cm,keepaspectratio]
   {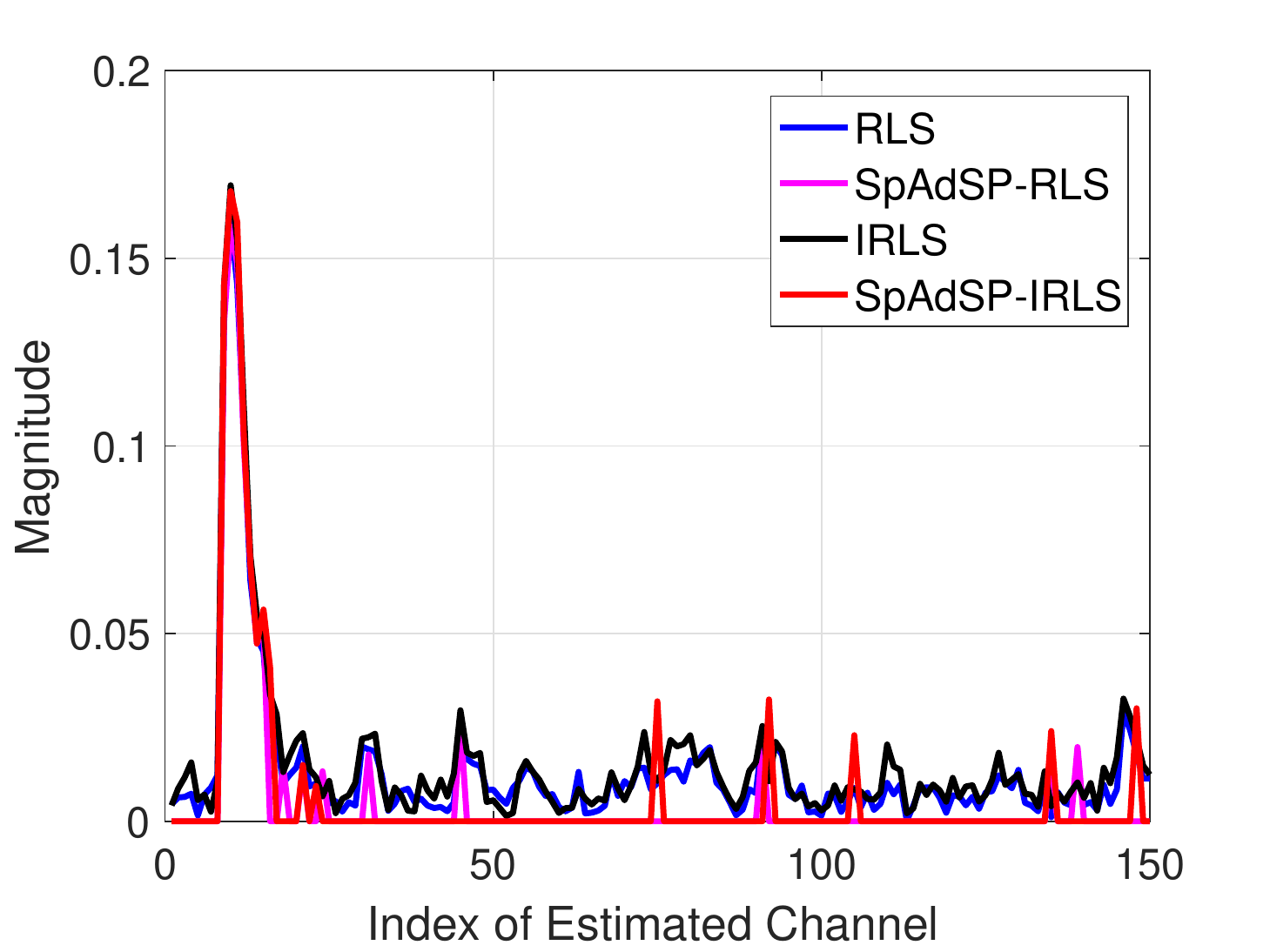}
   \caption{Demonstration of estimated channels by different algorithms.}
   \label{Figure3}
\end{figure}
\begin{figure}[!ht] 
   \hspace{1.1cm}
   \includegraphics[width=6.5cm,keepaspectratio]
   {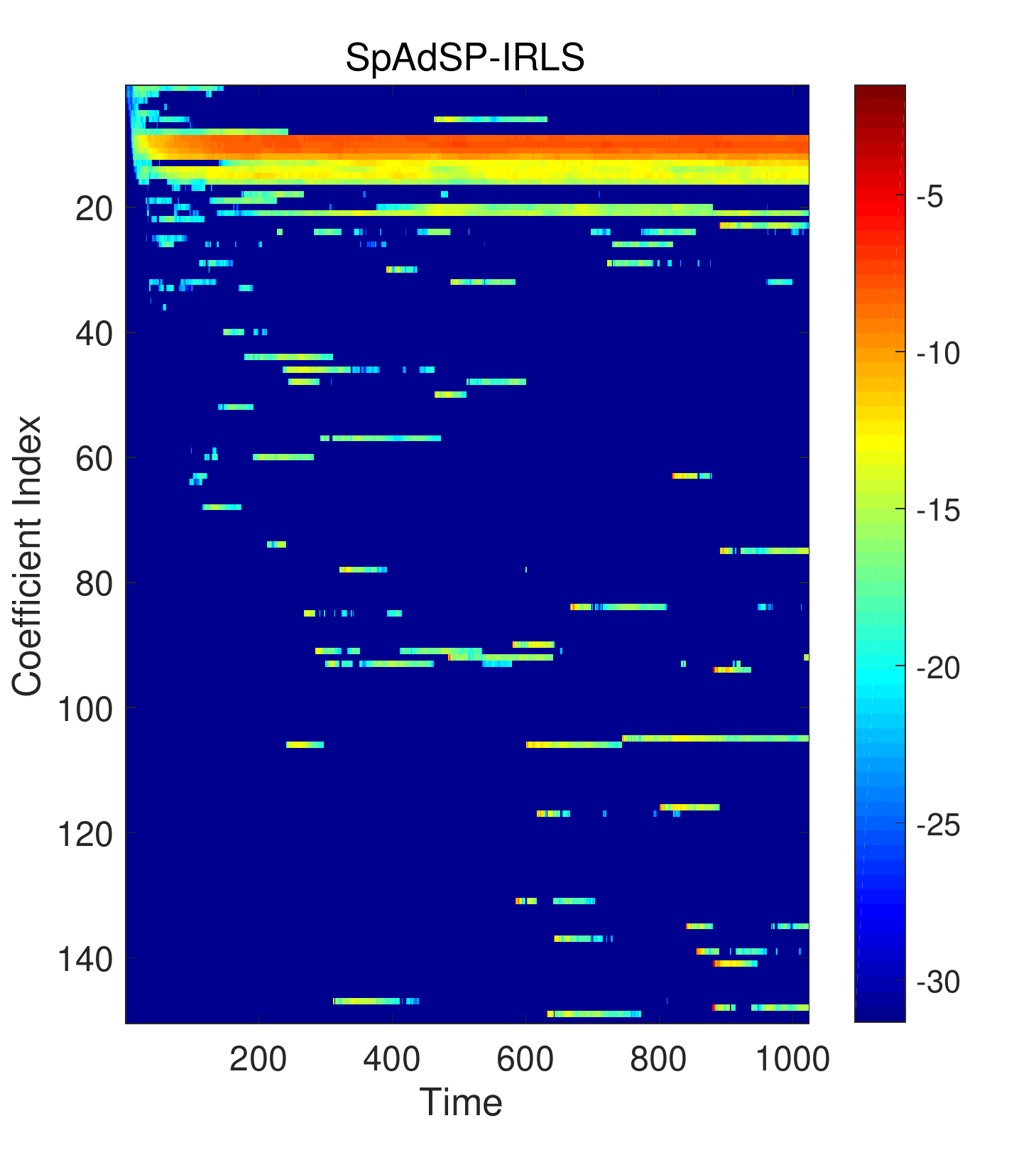}
   \caption{Demonstration of the sequence of SpAdSP-IRLS coefficient vector.}
   \label{Figure4}
\end{figure}
\begin{figure}[!ht] 
   \hspace{0.5cm}
   \includegraphics[width=7.5cm,keepaspectratio]
   {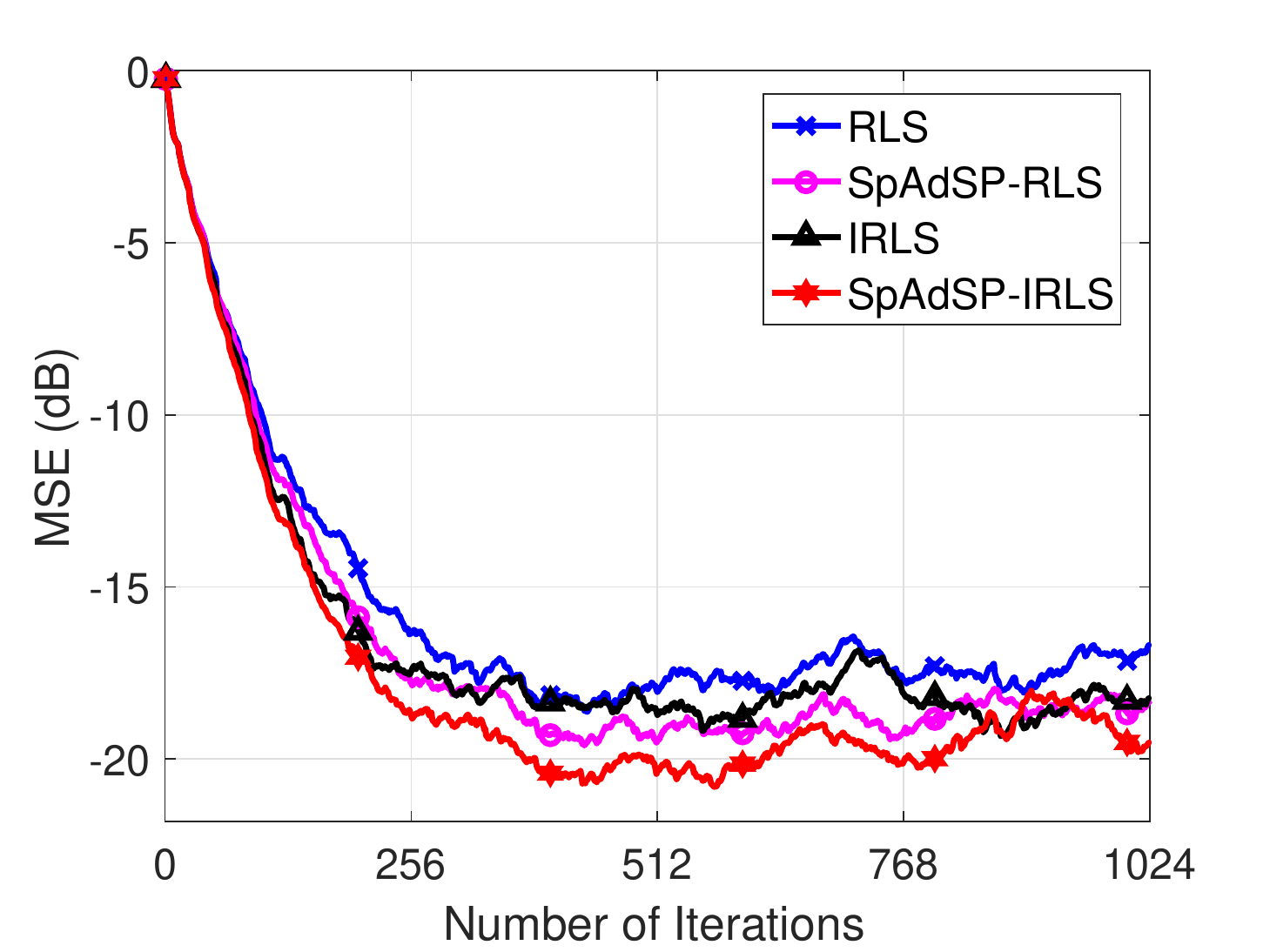}
   \caption{Performance comparison of different algorithms.}
   \label{Figure5}
\end{figure}
\begin{figure}[!ht] 
   \hspace{0.5cm}
   \includegraphics[width=7.5cm,keepaspectratio]
   {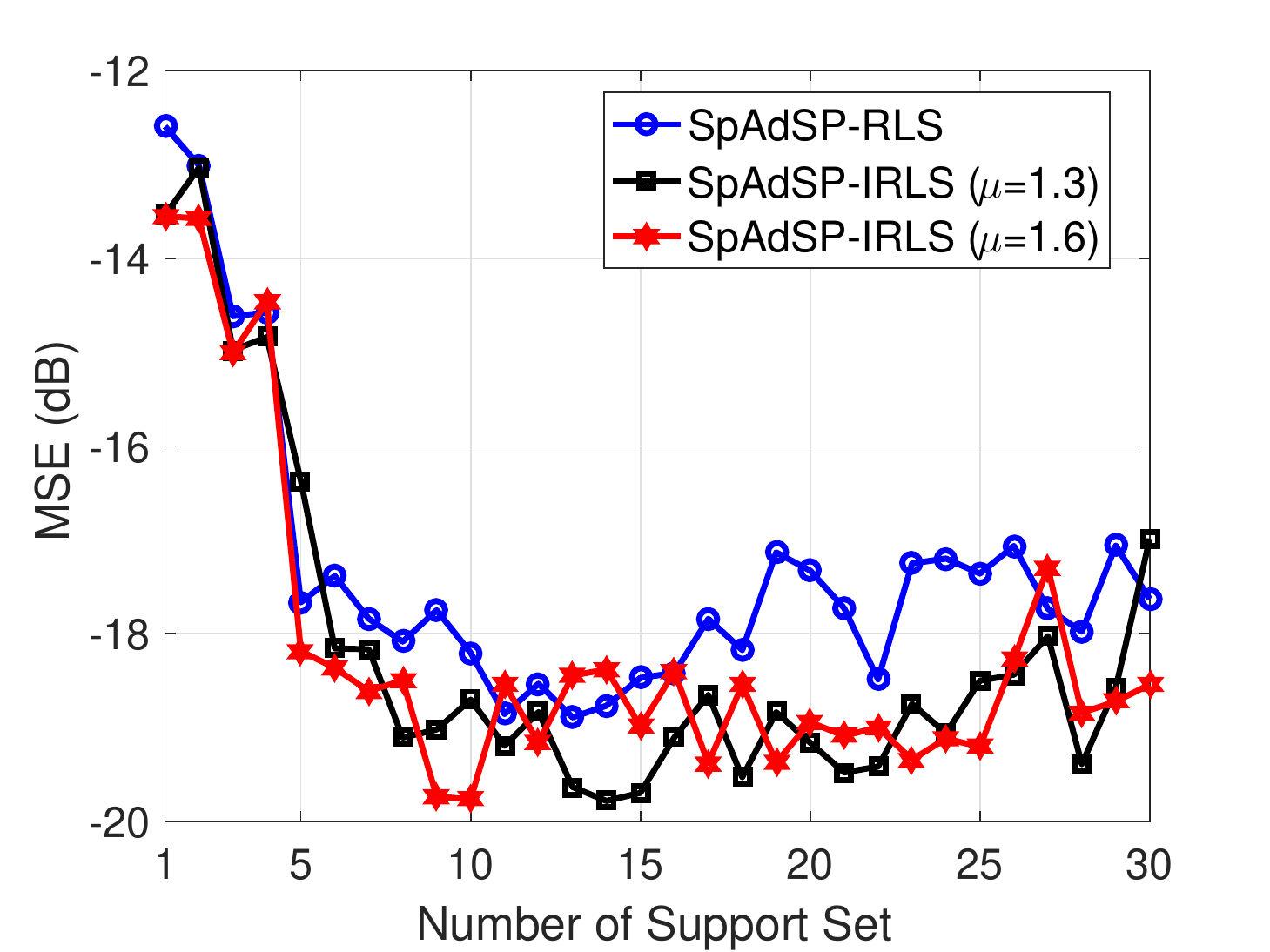}
   \caption{Steady-state Performance comparison of the SpAdSP-IRLS.}
   \label{Figure6}
\end{figure}

In Fig.~\ref{Figure3}, snapshots of estimated channels are shown.
We can notice that the channel shows the sparse structure and the SpAdSP-IRLS achieved the most sparse feature.
We also demonstrate that the position and magnitude of the SpAdSP-IRLS coefficients vary with time in Fig.~\ref{Figure4}.
In Fig.~\ref{Figure5}, the SpAdSP-IRLS obtained the best performance while only $15$ coefficients were employed, meaning that the complexity of it was obviously decreased.
To further explore the characteristic of the support set $s$, we measured steady-state error of the proposed SpAdSP-IRLS under different $\mu$.
From Fig.~\ref{Figure6}, $\mu$ introduced in the updating equation indeed improved the performance in different $s$.
In addition, while $s$ was chosen to be small, the obvious improvement of the SpAdSP-IRLS can be obtained due to exploitation of the sparse structure and noise suppression.
However, with the increase of $s$, there exists the performance loss because the performance of the SpAdSP-IRLS gradually approaches that of non-sparse counterpart.

\section{Conclusion and Outlook}
\label{Conclusion}

In this paper, a dynamic compressive sensing technique named the sparse adaptive subspace pursuit-improved recursive least squares (SpAdSP-IRLS) was proposed by only updating the support set in the IRLS.
Compared with the high complexity of the standard RLS algorithm, the SpAdSP-IRLS achieves a lower complexity while obtains a better performance.
The SpAdSP-IRLS was tested by synthetical and at-sea experimental data and shown to significantly outperform conventional RLS at reduced complexity.

An important area of focus for us in the future is to develop one DCS-based second-order algorithm with lower complexity. While SpAdSP-IRLS has successfully reduced the complexity of the standard RLS from $\mathcal{O}(3L^2+4L)$ to $\mathcal{O}(L^2+2L(s+1)+10s)$ for the length-$L$ channel with the size-$s$ support set, its complexity remains quadratic in terms of $L$. To achieve a super-linear convergence rate and linear complexity simultaneously, one option is to use fast RLS algorithms \cite{Cioffi84,Slock91} with the DCS technique. However, in practice, there are too many free parameters that require fine-tuning. Another approach is to use the diagonal quasi-Newton method \cite{bordes2009sgd,andrei2019diagonal,ma2020apollo} to achieve this objective. We will leave these investigations to future work.


\begin{thebibliography}{10}

\bibitem{cotter2002sparse}
S.~F. Cotter and B.~D. Rao, ``Sparse channel estimation via matching pursuit
  with application to equalization,'' {\em IEEE Trans. Commun.}, vol.~50,
  pp.~374--377, Mar. 2002.

\bibitem{Li07}
W.~Li and J.~C. Preisig, ``Estimation of rapidly time-varying sparse
  channels,'' {\em IEEE J. Ocean. Eng.}, vol.~32, pp.~927--939, Oct. 2007.

\bibitem{Berger09}
C.~R. Berger, S.~Zhou, J.~C. Preisig, and P.~Willett, ``Sparse channel
  estimation for multicarrier underwater acoustic communication: From subspace
  methods to compressed sensing,'' {\em IEEE Trans. Signal Process.}, vol.~58,
  pp.~1708--1721, Mar. 2009.

\bibitem{Pelekanakis13}
K.~Pelekanakis and M.~Chitre, ``New sparse adaptive algorithms based on the
  natural gradient and the ${L}_{0}$ -norm,'' {\em IEEE J. Ocean. Eng.},
  vol.~38, pp.~323--332, Apr. 2013.


\bibitem{Pelekanakis14}
K.~Pelekanakis and M.~Chitre, ``Adaptive sparse channel estimation under
  symmetric alpha-stable noise,'' {\em IEEE Trans. Wireless Commun.}, vol.~13,
  pp.~3183--3195, Jun. 2014.

\bibitem{Duan18}
W.~Duan, J.~Tao, and Y.~R. Zheng, ``Efficient adaptive turbo equalization for
  multiple-input-multiple-output underwater acoustic communications,'' {\em
  IEEE J. Ocean. Eng.}, vol.~43, pp.~792--804, Jul. 2018.

\bibitem{Tao19}
J.~Tao, Y.~Wu, X.~Han, and K.~Pelekanakis, ``Sparse direct adaptive
  equalization for single-carrier {MIMO} underwater acoustic communications,''
  {\em IEEE J. Ocean. Eng.}, vol.~45, pp.~1622--1631, Oct. 2020.

\bibitem{Chen09}
Y.~Chen, Y.~Gu, and A.~O. Hero, ``Sparse {LMS} for system identification,'' in
  {\em Proc. Int. Conf. Acoust., Speech, Signal Process. (ICASSP)},
  pp.~3125--3128, 2009.

\bibitem{Gu09}
Y.~Gu, J.~Jin, and S.~Mei, ``$l_0$ norm constraint {LMS} algorithm for sparse
  system estimation,'' {\em IEEE Signal Process. Lett.}, vol.~16, pp.~774--777,
  Sep. 2009.

\bibitem{Eksioglu111}
E.~M. Eksioglu and A.~K. Tanc, ``{RLS} algorithm with convex regularization,''
  {\em IEEE Signal Process. Lett.}, vol.~18, pp.~470--473, Aug. 2011.

\bibitem{Qin19}
Z.~Qin, J.~Tao, X.~Wang, X.~Luo, and X.~Han, ``Direct adaptive equalization
  based on fast sparse recursive least squares algorithms for multiple-input
  multiple-output underwater acoustic communications,'' {\em J. Acoust. Soc.
  Amer.}, vol.~145, pp.~277--283, Apr. 2019.

\bibitem{Duttweiler00}
D.~L. Duttweiler, ``Proportionate normalized least-mean-squares adaptation in
  echo cancelers,'' {\em IEEE Trans. Audio, Speech, Lang. Process.}, vol.~8,
  pp.~508--518, Sep. 2000.

\bibitem{Benesty02}
J.~Benesty and S.~L. Gay, ``An improved {PNLMS} algorithm,'' in {\em Proc. Int.
  Conf. Acoust., Speech, Signal Process. (ICASSP)}, (Orlando, FL, USA),
  pp.~1881--1884, 2002.

\bibitem{Loganathan09}
P.~Loganathan, A.~W.~H. Khong, and P.~A. Naylor, ``A class of
  sparseness-controlled algorithms for echo cancellation,'' {\em IEEE Trans.
  Speech Audio Process.}, vol.~17, pp.~1591--1601, Nov. 2009.

\bibitem{Qin20}
Z.~Qin, J.~Tao, and Y.~Xia, ``A proportionate recursive least squares algorithm
  and its performance analysis,'' {\em IEEE Trans. Circuits Syst. II, Exp.
  Briefs.}, vol.~68, pp.~506--510, Jan. 2021.

\bibitem{QinDSP22}
Z.~Qin, J.~Tao, Y.~Xia, and L.~Yang, ``Proportionate $\text{RLS}$ with $l_1$
  norm regularization: Performance analysis and fast implementation,'' {\em
  Digital Signal Process.}, vol.~122, p.~103366, Apr. 2022.

\bibitem{Pelekanakis10}
K.~Pelekanakis and M.~Chitre, ``Comparison of sparse adaptive filters for
  underwater acoustic channel equalization/estimation,'' in {\em Proc. IEEE
  Int. Conf. Commun. Syst.}, pp.~395--399, Nov. 2010.

\bibitem{Wu13}
F.~Wu, Y.~Zhou, F.~Tong, and R.~Kastner, ``Simplified p-norm-like constraint
  lms algorithm for efficient estimation of underwater acoustic channels,''
  {\em J. Marine Sci. Appl.}, vol.~12, pp.~228--234, Jun. 2013.

\bibitem{Tian19}
Y.~Tian, X.~Han, J.~Yin, Q.~Liu, and L.~Li, ``Group sparse underwater acoustic
  channel estimation with impulsive noise: Simulation results based on arctic
  ice cracking noise,'' {\em J. Acoust. Soc. Amer.}, vol.~146, pp.~2482--2491,
  Oct. 2019.

\bibitem{Vaswani08}
N.~Vaswani, ``Kalman filtered compressed sensing,'' in {\em Proc. 15th IEEE
  Intl. Conf. Image Proc. (ICIP)}, pp.~893--896, 2008.

\bibitem{Ziniel13}
J.~Ziniel and P.~Schniter, ``Dynamic compressive sensing of time-varying
  signals via approximate message passing,'' {\em IEEE Trans. Signal Process.},
  vol.~61, pp.~5270--5284, Nov. 2013.

\bibitem{Mileounis10}
G.~Mileounis, B.~Babadi, and N.~Kalouptsidis, ``An adaptive greedy algorithm
  with application to nonlinear communications,'' {\em IEEE Trans. Signal
  Process.}, vol.~58, pp.~2998--3007, Jun. 2010.

\bibitem{Vlachos12}
E.~Vlachos, A.~S. Lalos, and K.~Berberidis, ``Stochastic gradient pursuit for
  adaptive equalization of sparse multipath channels,'' {\em IEEE J. Emerg.
  Sel. Topics Circuits Syst.}, vol.~2, pp.~413--423, Sep. 2012.

\bibitem{Zachariah12}
D.~Zachariah, S.~Chatterjee, and M.~Jansson, ``Dynamic iterative pursuit,''
  {\em IEEE Trans. Signal Process.}, vol.~60, pp.~4967--4972, Sep. 2012.

\bibitem{Qin202}
Z.~Qin, J.~Tao, and X.~Han, ``Dynamic compressive sensing based adaptive
  equalization for underwater acoustic communications,'' in {\em Proc. MTS/IEEE
  Global OCEANS Conf.}, pp.~1-- 5, Oct. 2020.

\bibitem{qin2022adaptive}
Z.~Qin, J.~Tao, F.~Qu, and Y.~Qiao, ``Adaptive equalization based on dynamic
  compressive sensing for single-carrier multiple-input multiple-output
  underwater acoustic communications,'' {\em J. Acoust. Soc. Amer.}, vol.~151,
  pp.~2877--2884, May 2022.

\bibitem{TaoOceans19}
J.~Tao, Y.~Wu, Q.~Wu, and X.~Han, ``Kalman filter based equalization for
  underwater acoustic communications,'' in {\em Proc. MTS/IEEE Global OCEANS
  Conf.}, pp.~1-- 5, Jun. 2019.

\bibitem{stojanovic94}
M.~Stojanovic, J.~Catipovic, and J.~Proakis, ``Phase-coherent digital
  communicaitons for underwater acoustic channels,'' {\em IEEE J. Ocean. Eng.},
  vol.~19, pp.~100--111, Jan. 1994.

\bibitem{Dai09}
W.~Dai and O.~Milenkovic, ``Subspace pursuit for compressive sensing signal
  reconstruction,'' {\em IEEE Trans. Inf. Theory}, vol.~55, pp.~2230 -- 2249,
  May 2009.

\bibitem{Cioffi84}
J.~Cioffi and T.~Kailath, ``Fast, recursive-least-squares transversal filters
  for adaptive filtering,'' {\em IEEE Trans. Speech Audio Process.}, vol.~32,
  pp.~304--337, Apr. 1984.

\bibitem{Slock91}
D.~T.~M. Slock and T.~Kailath, ``Numerically stable fast transversal filters
  for recursive least squares adaptive filtering,'' {\em IEEE Trans. Signal
  Process.}, vol.~39, pp.~92--114, Jan. 1991.

\bibitem{bordes2009sgd}
A.~Bordes, L.~Bottou, and P.~Gallinari, ``{SGD}-{QN}: Careful quasi-newton
  stochastic gradient descent,'' {\em J. Mach. Learn. Res.}, vol.~10,
  pp.~1737--1754, 2009.

\bibitem{andrei2019diagonal}
N.~Andrei, ``A diagonal quasi-newton updating method for unconstrained
  optimization,'' {\em Numerical Algorithms}, vol.~81, no.~2, pp.~575--590,
  2019.

\bibitem{ma2020apollo}
X.~Ma, ``Apollo: An adaptive parameter-wise diagonal quasi-newton method for
  nonconvex stochastic optimization,'' {\em arXiv preprint arXiv:2009.13586},
  2020.

\end{thebibliography}

\end{document}